# EPB-TBM tunnel under internal pressure: Assessment of serviceability

N.A. Labanda & A.O. Sfriso
*SRK Consulting, Argentina*

D. Tsingas & R. Aradas
*Jacobs, Argentina*

M. Martini
*Salini-Impregilo S.p.A., Italy*

ABSTRACT: The Mantanza-Riachuelo basin recovery is one of the most ambitious environmental projects under construction in Argentina. In this context, the sanitary bureau of the metropolitan area of Buenos Aires (AySA) is building a sewage collection network to transport the waste water of the population in the southern area of the city, composed by almost five million people. The most complex tunnel in this big project is named *Lot 3*, an outfall EPB-TBM tunnel starting at a shaft located at the *Rio de la Plata* margin and running under the river 12 km to a discharge area.
The tunnel runs through soft clay belonging to the *post-pampeano* formation and dense sands of the *Puelchese* formation. In operation, it will be pressurized by a pumping station which will produce a piezometer head that, in the first 2000 m, might be eventually higher than the confining pressure around the tunnel.
This paper presents the numerical analysis of the structural forces acting on the tunnel rings using a risk-oriented approach that considers the stochastic nature of materials, stratigraphy and tunnel-ground interaction. The compression of the lining is evaluated and compared with field measurements in order to predict the structural forces and the risk of the rings going into tension beyond the structural capacity of the system.

## 1 INTRODUCTION

The *Matanza-Riachuelo river* flows along the riparian lands between the City and the Province of Buenos Aires (Argentina). It is a water stream 64 km long that flows into the *Rio de la Plata*. The river is heavily contaminated with industrial and residential wastewater discharged from both margins with limited -if any- pre-discharge water treatment process.

The *Matanza-Riachuelo basin recovery project* is an ambitious plan to manage the wastewater from the left margin of the *Rio de la Plata* through a sewerage interceptor tunnel, a treatment plant and a discharge tunnel into the *Rio de la Plata*. The project is divided into several contracts. One of them, named *Lot 3*, holds the outfall tunnel, an EPB-TBM tunnel running 35 meters below the riverbed of the *Rio de la Plata*. A global scheme of the outfall, which is the subject of this paper, is presented in Figure 1. The tunnel has an internal diameter of 4.3 m and a total length of 12 km, including a 1.5 km diffusor zone with 34 standpipe risers, 28 m long, that daylight in the riverbed. It is being driven through soft clays of the *Postpampeano* formation and dense sands from the *Puelchense* formation. Water flows by gravity from a pumping station onshore built into the access shaft. Figure 2 shows the longitudinal elevation of the tunnel proposed in pre-feasibility stage.

The tunnel will be subjected to a maximum internal pressure of 550 kPa and an external water pressure from 330 to 420 kPa, resulting in a net outwards pressure 130 to 220 kPa. Further information about the tunnel and its design can be found in Aradas et al. (2019). EPB-TBMs employ muck to balance water and soil pressure at the tunnel face. By managing the excavation parameters i.e., screw conveyor speed, advance rate, thrust force, face support pressure, etc., the ground relaxation and, consequently, structural forces in the lining can be somewhat controlled. This aspect, usually not of high interest, is important in the case of segmental pressurized tunnels working at high internal pressure.

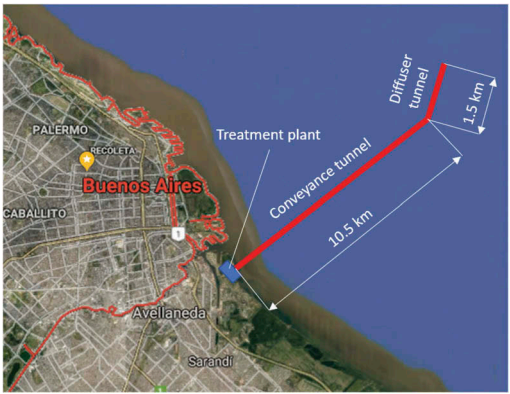

Figure 1. Location of Matanza-Riachuelo outfall (Buenos Aires, Argentina). The thin red line is the border between the city of buenos aires and the province of buenos aires. The Riachuelo river is along the south border.

A common feature found when reviewing case histories of pressurized tunnels in soft ground is the recommendation to neglect the contribution of the external ground confinement as a safety redundancy. This hypothesis, despite being conservative from the perspective of confinement, produces an unrealistic estimate of the deformation of the lining ring and therefore of the resulting bending moments. On the other hand, the adoption of a nonzero effective external ground pressure is challenging as it is largely dependent on the TBM-soil interaction during excavation and on the long-term creep behavior of the soil.

An assessment of the compression induced in the segmental rings by ground pressure via stochastic numerical models and calibrated using back-analysis after field measurements is presented in this paper. Two rings are studied in this proposal: $N°237$ and $N°497$, embedded in soft clays and dense sands, respectively.

## 2 CONSTITUTIVE MODELING FOR SOIL AND LINING CONCRETE

The tunnel was designed in pre-feasibility stage to be driven through dense sands, as shown in Figure 2, where yellow layers represent the *Puelchense* formation. However, field conditions departed from this assumption in several places. For instance, high-plasticity clays belonging to the *Paranaense* formation were detected in the conveyor belt (Figure 3) betwen chainages PK 0+294 to PK 0+586. Ring $N°237$ placed at PK 0+316 and consequently, embedded in clay, and Ring $N°497$ placed at PK 0+681, embedded in dense sand, were instrumented with strain gauges whose results allow for evaluating the compression load of the tunnel lining in two different grounds.

The *Hardening soil small* (HSsmall) constitutive model was used to simulate the mechanical behavior of soils and calibrated by in situ and lab tests and local experience. Statistical variability of the friction angle was estimated using the 6-sigma method. A correlation between the compression index $C_c$ and the liquid limit $\omega_L$ for *Postpampeano* clays was first reported by (Sfriso 1997) with a $COV = 0.30$, and was later updated by (Ledesma 2008) as presented in Figure 4, with a $COV = 0.12$. Swelling index $C_s$, relevant for

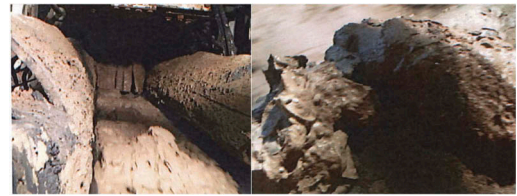

Figure 3. High-plasticity clay detected in the TBM's conveyor belt.

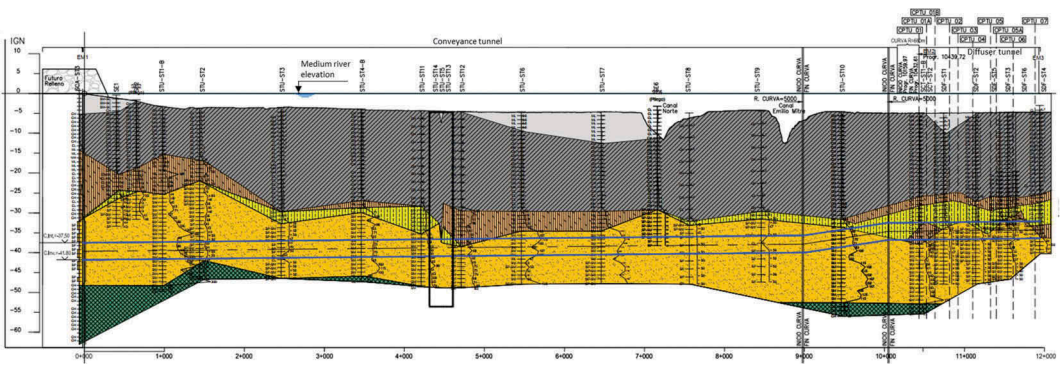

Figure 2. Matanza-Riachuelo outfall: longitudinal elevation of the tunnel proposed in pre-feasibility stage.

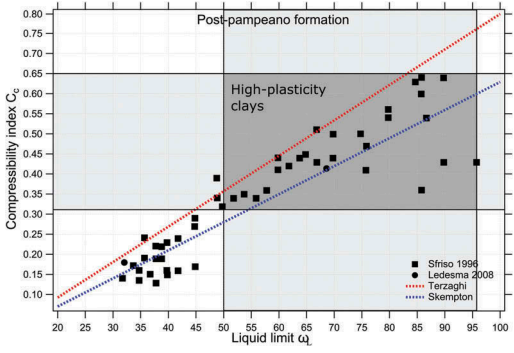

Figure 4. Compression index $C_c$ versus liquid limit for the post-pampeano formation.

Table 1. Constitutive models and parameters used.

| | Unit | Soft clay | Sandy soils |
|---|---|---|---|
| Model | - | HSsmall | HSsmall |
| Drainage | - | Undrained (A) | Drained |
| $\gamma$ | $kN/m^3$ | 16.0 | 20.5 |
| $\phi'$ | ° | P.D. | 32 |
| $c'$ | $kPa$ | 0 | 0 |
| $\psi$ | ° | 0 | 0 |
| $G_0^{ref}$ | $MPa$ | P.D. | 200 |
| $\gamma_{0.7}$ | - | $10^{-3}$ | $10^{-4}$ |
| $E_{ur}^{ref}$ | $MPa$ | P.D. | 120 |
| $E_{50}^{ref}$ | $MPa$ | P.D. | 40 |
| $E_{oed}^{ref}$ | $MPa$ | P.D. | 40 |
| $m$ | - | 1.0 | 0.5 |
| $\nu_{ur}$ | - | 0.20 | 0.20 |
| OCR | - | 1.00 | 1.00 |
| $K_0^{nc}$ | - | $1 - \sin \phi'$ | $1 - \sin \phi'$ |
| $R_{inter}$ | - | 0.90 | 0.70 |
| $k$ | $10^{-6}$ m/s | 0.005 | 1.00 |

the estimation of the interaction forces, correlates with $C_c$ by:

$$C_s = \frac{1}{5} | \frac{1}{10} C_c \quad and \quad C_c = 0.009(\omega_L - 10) \pm 0.12 \quad (1)$$

HSsmall requires the definition of elastic and hardening parameters $E_{oed}^{ref}$, $E_{ur}^{ref}$, $E_{50}^{ref}$ and $G_0^{ref}$, among other parameters. They were computed as

$$E_{oed}^{ref} = \frac{2.3(1 + e_0)p_{ref}}{C_c}, \quad (2)$$

$$E_{ur}^{ref} = \frac{2.3(1 + e_0)(1 + \nu_{ur})(1 - 2\nu_{ur})p_{ref}}{(1 - \nu_{ur})K_0 C_s}, \quad (3)$$

$$E_{50}^{ref} = 1.25 | 1.90 E_{oed}^{ref}, \quad (4)$$

$$G_0^{ref} = \frac{5}{2(1 + \nu)} E_{ur}^{ref}, \quad (5)$$

being $\nu_{ur}$ the Poisson modulus, $K_0$ the at rest earth pressure coefficient, $p_{ref}$ the reference pressure and $e_0$ the initial void ratio.

Uncertainties in the properties of the clay where the ring $N°237$ is placed were dealt with by considering a stochastic approach to bracket the prediction of the structural forces acting on the lining. The parameters employed are presented in Table 1, where some properties are defined using probability distributions that will be described in the next section. A deterministic set of parameters were used for dense sand because its influence is negligible for the ring under analysis.

The structural stiffness of the lining was reduced to account for the segment joints using the Muir-Wood expression (Muir Wood 1975)

$$I_{red} = I_{real} \left( \frac{4}{n} \right)^2, \quad (6)$$

where $I_{red}$ is the reduced moment of inertia, $I_{real}$ is the real moment of inertia and $n$ is the number of lining segments in each ring.

## 3 STATISTICAL CHARACTERIZATION

### 3.1 Soil parameters

The proposed risk analysis requires that the variability of soil parameters be defined together with the boundaries presented in Equation (4). A statistical characterization for the effective friction angle $\phi'$ and the compression index $C_c$ is presented in Figure 5. Probability distributions were obtained using the experimental data shown in Figure 4 and classical correlations for the friction angle. Considering values $\bullet \pm \sigma_\bullet$ for parameters $C_c$, $\phi'$ and boundary multipliers in Equations (1) and (4), $2^4 = 16$ permutations can be performed to reproduce drained triaxial tests, where $\bullet$ is the mean value and $\sigma_\bullet$ is the standard deviation of the considered parameter. Results are presented in Figure 6 and compared with three experimental tests for this project. Good fitting is obtained for consolidation mean pressures of $p = 50 kPa$, $p = 100 kPa$ and $p = 200 kPa$.

### 3.2 Lining segment concrete

The Young's modulus of concrete is required to compute stresses and forces out of microstrains measured by the monitoring system. A histogram of 480 simple compression tests, a normal probability function and a cumulative probability function are shown in Figure 7. In order to compute the concrete stiffness at low strains $E_{ci}$, the following expression is used (CEB-FIP 1993)

$$E_{ci} = E_{co}\left[\frac{f_{cm}}{f_{cmo}}\right]^{1/3}, \quad (7)$$

with $E_{co} = 21500 MPa$, $f_{cmo} = 10 MPa$ and $f_{cm}$ unconfined compression resistance of the sample.

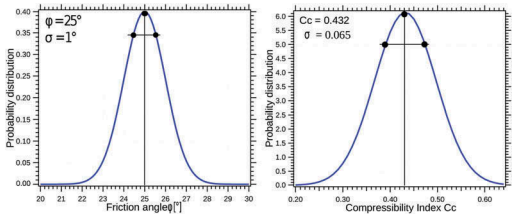

Figure 5. Probability distribution for friction angle and compressibility index of the Postpampeano formation.

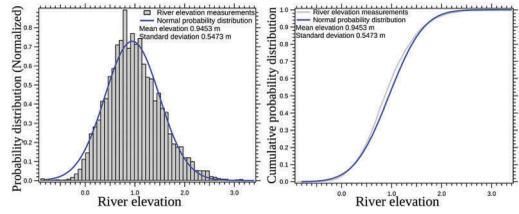

Figure 8. Water level in the *Rio de la Plata*. Probability distribution and cumulative distribution.

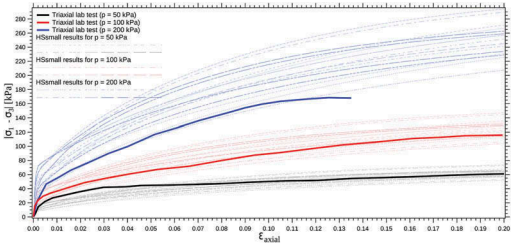

Figure 6. Drained triaxial test, comparisons between laboratory tests and HSsmall simulations.

Considering a mean value $f_{cm} = 59.6 MPa$ and a standard deviation $\sigma_{f_{cm}} = 4.94 MPa$, deformations measured by strain gauges can be translated into structural forces.

### 3.3 Water level in the river

Variable water level in the river is one of the key contributors to uncertainty of the compression forces acting on the tunnel lining, making its statistical characterization a critical aspect of the analysis. The Argentine Naval Hydrographic Service provided the time-history evolution of the river over the last few years. Taking the data corresponding to the first semester of 2018, where the rings under analysis were installed, a normal probability function is fitted and presented in Figure 8, and compared with field measurements. A mean elevation 0.945 $m$ and a standard deviation 0.547 $m$ were considered in the analyses.

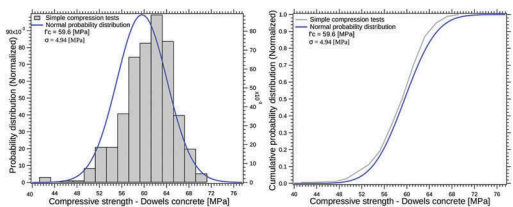

Figure 7. Histogram of unconfined compression test results for 480 samples of lining concrete. Probability distribution and cumulative distribution.

## 4 NUMERICAL RESULTS AND COMPARISON WITH FIELD MEASUREMENTS

### 4.1 *Ground relaxation estimate by 3D modeling in ring N°237*

A three dimensional model of the excavation was developed to estimate the ground relaxation induced by the TBM drive in ring $N°237$. The TBM shield was simulated considering a face contraction $c_{ref} = 0.4\%$ and a tail-to-face contraction increment $c_{inc,axial} = 0.03571\%/m$. A grout pressure equal to the total field stress plus 0.5 bar was applied in the TBMs tail. The face pressure was calibrated with data provided by the sensors in the TBM, imposing a normal surface load equal to the measured value presented in Figure 9. The 3D model is presented in Figure 10. Due to the high computational effort required by this kind of simulations, mean values for all materials were used and a single model is analyzed. The excavation sequence was repeated until a reasonably stabilized zone was obtained.

Settlements in the tunnel crown and effective vertical stresses in the stabilized section are plotted in Figure 11. The mean value for settlement is close to 30 mm, while the effective vertical stress is around 145 kPa. These values were used to fit a ground relaxation factor in the two dimensional models presented in the next section.

### 4.2 *Structural forces in ring N°237*

In order to evaluate the risk of tension forces being developed in the ring, a 2D model was developed and the influence of uncertainties in input data in

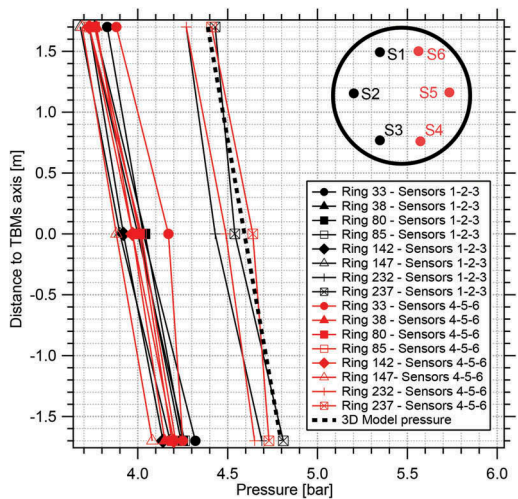

Figure 9. Sensor data from the TBM face.

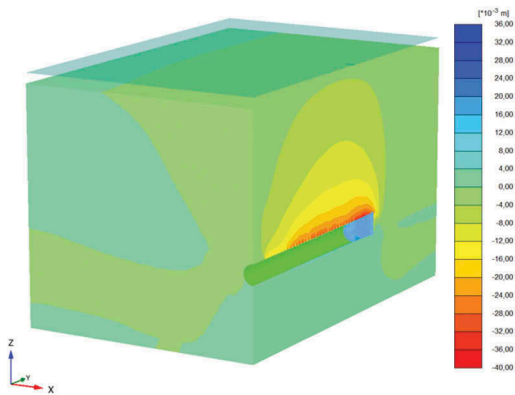

Figure 10. 3D model of settlements after excavation.

Equations (1) and (4), river water level and concrete quality were studied.

The numerical model and construction stages are summarized in Figure 12. The excavation sequence was simulated in 2D by partial stress relaxation using the $\beta$-method, i.e. applying $\Sigma Mstage = 1 - \beta < 1$. Grout pressure in TBMs tail was simulated as a distributed load, the (impervious) lining was activated and consolidation until 99% dissipation of excess pore pressure -the expected condition by the time the tunnel starts operating- was allowed for. In operation, an internal head of 15.5 m, 14.0 to 15.1 meters above the river level was applied. $2^5 = 32$ models were generated by performing all permutations of data presented in Table 2.

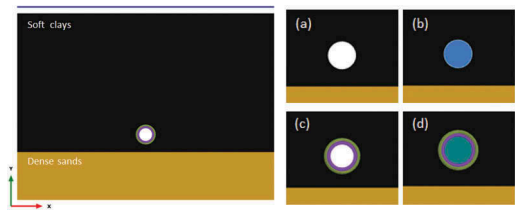

Figure 12. 2D model of ring $N°237$. (a) Excavation using $\Sigma M\, stage < 1$ (b) grout pressure (c) Activation of lining and consolidation (d) Internal pressure in operation.

Table 2. Parameter ranges used to compute ground relaxation curves.

|  | Unit | Lower bound | Upper Bound |
| --- | --- | --- | --- |
| $C_c$ | - | 0.367 | 0.497 |
| $C_s$ (eq. 1) | - | 0.1 | 0.2 |
| $E_{50}^{ref}$ (eq. 4) | - | 1.25 | 1.90 |
| $\phi'$ | ° | 24 | 26 |
| River Elevation | m | 0.398 | 1.492 |

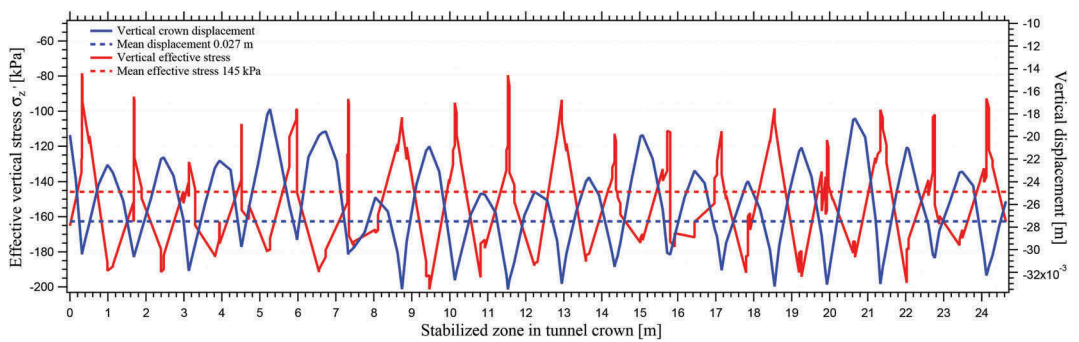

Figure 11. Settlements of tunnel crown and effective vertical stress in stabilized section.

Figure 13 shows the obtained ground relaxation curves and both displacement and effective vertical stress measured in the tunnel crown, plotted in terms of $\Sigma M\ stage$. The purpose of these curves is to obtain, in a 2D model, a stress state similar to the stabilized section of a 3D model.

By getting into the graph with the mean crown displacement obtained in the 3D model, and intersecting curves corresponding to the 2D crown displacement (in blue), a range of $\Sigma M\ stage$ values were obtained. The range of effective contact stresses obtained reasonably contains the mean value obtained in the 3D model. Values adopted for the analysis are $\Sigma M\ stage = [0.35, 0.40, 0.45, 0.50, 0.55, 0.60]$. Tables 2 and 3 show values employed to calculate structural forces in the lining. Together with the $\Sigma M\ stage$ values, a set of $6 \times 2^6 = 384$ numerical models were obtained.

Figure 14 plots translation and ovalization components of the tunnel lining obtained with all permutations, during construction. The model shows that, for low relaxation values, a little rebound of the tunnel is observed and an ovalization pattern where the vertical stress is larger than the horizontal stress is obtained. While the ground relaxation increases, a tunnel settlement and an invertion in the tunnel ovalization is obtained. Figure 15 shows the obtained structural forces for all models in the construction stage, comparing those with field data measured in ring $N°237$ using strain gauges. The actual concrete stiffness was used to translate deformations into forces, as indicated with markers for the mean value and lines for the standard deviation band.

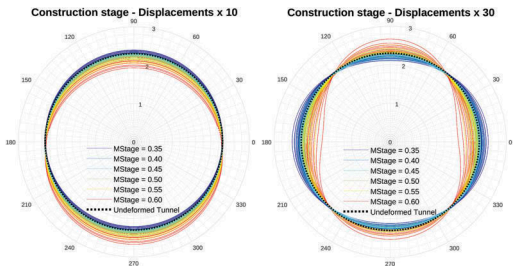

Figure 14. Translation and ovalization of ring $N°237$ during construction.

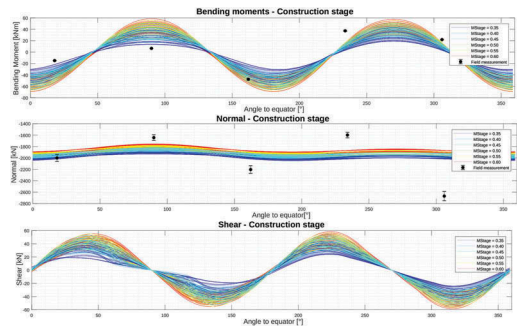

Figure 15. Structural forces of ring $N°237$ in construction stage and comparison with field measurements.

It can be seen that the proposed procedure prodcues a good fitting with field data for both bending moments and normal forces. For low $\Sigma M\ stage$ values, hihger normal forces and lower bending moments are obtained, being the last in agreement with the ovalization pattern. If $\Sigma M\ stage$ is higher, normal forces decrease and bending moments increase. Figure 16 shows a comparison of the mobilized shear stress $\tau_{mob}$ for a 2D model and a stabilized section of the 3D model. For sake of simplicity, a single representative 2D model is shown but similar stress patterns are obtained in all permutations. It is interesting to see that the mobilized shear, a measure of the tunnel arching effect, is similar in both cases. After ensuring the validation of the proposed methodology, a study of the tunnel lining in operation was performed and the risk of tension forces being developed was analyzed.

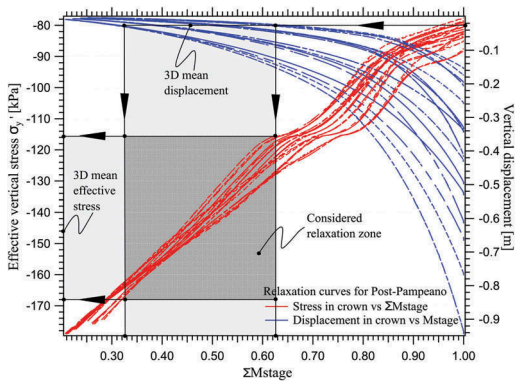

Figure 13. Ground relaxation curves for $2^5 = 32$ permutations in soil parameters and river water level. Ring $N°237$.

Table 3. Concrete parameters used in permutations to compute structural forces.

|  | Unit | Lower bound | Upper Bound |
| --- | --- | --- | --- |
| $f_{cm}$ | MPa | 54.6 | 64.5 |

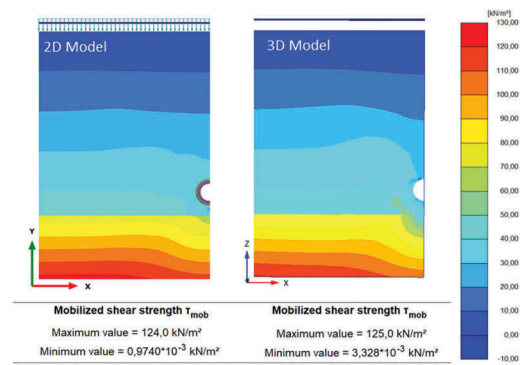

Figure 16. Comparison of $\tau_{mob}$ between the 2D model and a stabilized section of the 3D model.

### 4.3 Structural forces in operation in ring $N°237$

Lining forces in operation in ring $N°237$ are evaluated in this section. Figure 17 shows the translation and ovalization for the tunnel under internal pressure. Results are similar than those obtained for the construction stage, with a small increment in tunnel settlements due to the waste water weight. Ovalization remains almost equal by virtue of the hydrostatic not generating higher deformations compared with the construction stage. Figure 18 plots the structural forces for all considered permutations. Results show that bending moments and shear forces are almost the same than during construction. Normal compression forces, however, decrease dramatically from values $-1900 \pm 100 kN$ before the internal pressure to $-200 \pm 150 kN$ in service. Despite this fact, even for high relaxation values i.e. $\Sigma M\ stage = 0.60$, lining is still under compression, and tension in connecting bolts is avoided.

### 4.4 Structural forces in ring $N°497$

Structural forces and deformations in the most adverse chainage of the tunnel, represented by ring $N°497$, is presented in this section.

The procedure for the back analysis used in this case is different than the used in the previous case, avoiding 3D calculations and fixing the ground relaxation by means of $\Sigma M\ stage$ and field measurements. The simulation procedure is the same than in the previous case. Soil properties for the soft clay and overlying materials were defined using the mean values presented in previous sections, while parameters for dense sand were considered to be permuted using bounds presented in Table 4.

A two dimensional base model considering the stratigraphy corresponding to chainage PK 0 +681 is presented in Figure 19. The construction procedure is the same than in the previous case.

The effective friction angle $\phi'$ was defined as the sum of the constant volume friction angle $\phi_{cv} = 30°$ and the dilatancy angle $\psi$ presented above.

Ground relaxation curves for current stratigraphy are shown in Figure 20. Displacement are considerably lower than those observed for soft clays in Figure 13. Also, a lower ground mobilization is required to achieve the same confinement pressure on the lining. Concrete parameters presented in Table 3, together with relaxations $\Sigma M\ stage = [0.35, 0.45, 0.55]$, produce 192 sets of paramteres that were employed to do the stochastic analysis. Results are expressed in terms of bending moments, normal and shear forces during construction, and are shown together with field measurements for ring $N°497$ in Figure 21. Good fitting is found between numerical and experimental results in bending moments, while predictions in normal forces tend to underestimate the compression forces in the lining. Tunnel translation and ovalization during construction are plotted in Figure 22, where vertical displacements are negligible while ovalization tends to increase with ground relaxation.

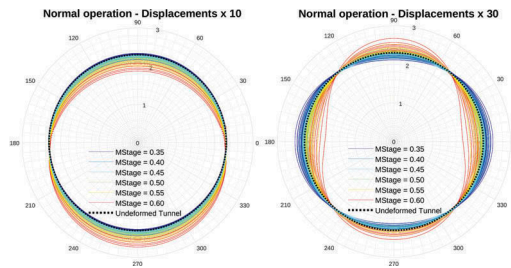

Figure 17. Translation and ovalization of ring $N°237$ in normal operation stage.

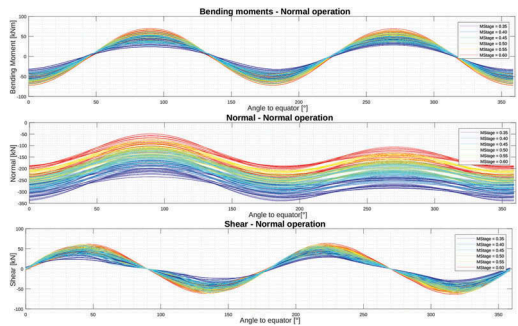

Figure 18. Structural forces of ring $N°237$ in normal operation.

Table 4. Parameters used to compute ground relaxation curves in dense sands.

|  | Unit | Lower bound | Upper Bound |
|---|---|---|---|
| $\psi$ | ° | 8.0 | 10.0 |
| $E_{50}^{ref}$ | [kPa] | 82000 | 98000 |
| $E_{ur}^{ref}$ | [kPa] | 200000 | 240000 |
| $G_0^{ref}$ | [kPa] | 280000 | 400000 |
| River Elevation | m | 0.398 | 1.492 |

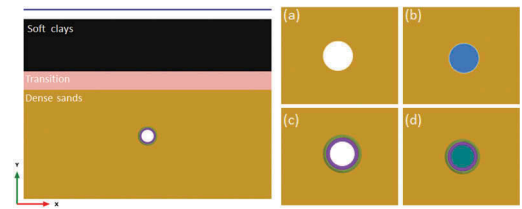

Figure 19. 2D base model for ring $N°497$. (a) Excavaction with $\Sigma M\ stage$ (b) grout pressure application (c) Lining construction (d) Tunnel under internal pressure in normal operation.

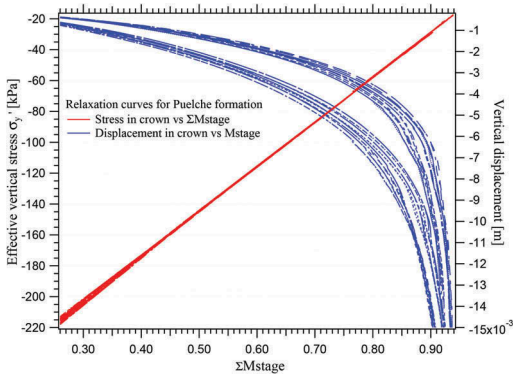

Figure 20. Ground relaxation curves for $2^5 = 32$ permutations in soil parameters and river water level. Ring $N°497$.

### 4.5 Structural forces in operation, ring $N°497$

Using the calibrated numerical model, an assessment of serviceability in operation (under internal pressure) is presented in this section.

Bending moments, normal and shear forces in the lining are presented in Figure 23, while translation and ovalization are shown in Figure 24.

Similar to ring $N°237$, internal pressure in the tunnel does not change much the bending moments, shear forces and lining deformations. However, normal forces were affected, being the internal pressure produced by the waste water transportation similar or even higher than the ground and water pressure applied in the tunnel lining. In this aspect, the behavior is different than the section embedded in clay; dense sands produce less pre-compression in the lining, thus increasing the probability to experience tensile forces in the joint bolts.

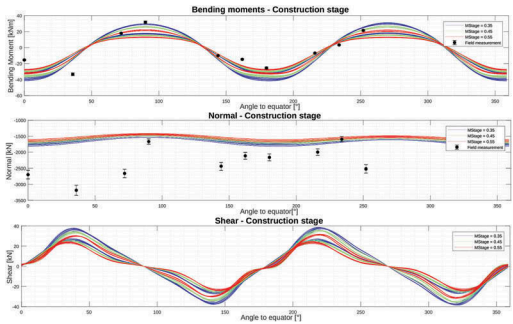

Figure 21. Structural forces of ring $N°497$ in construction stage and comparisons with field measurements.

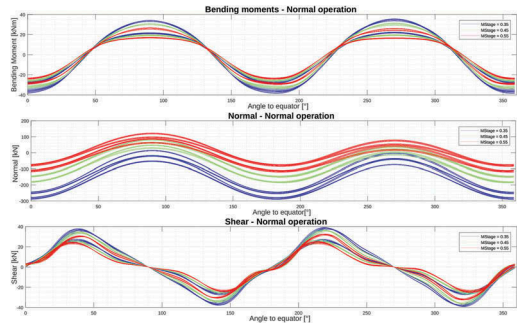

Figure 23. Structural forces of ring $N°497$ in operation.

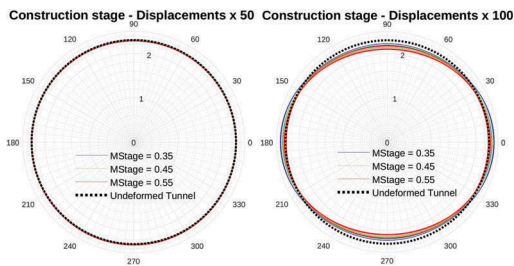

Figure 22. Translation and ovalization of ring $N°497$ during construction.

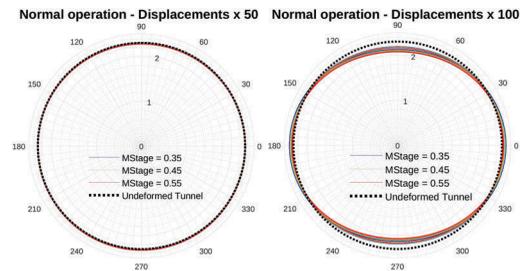

Figure 24. Translation and ovalization of ring $N°497$ in operation.

## 5 CONCLUSIONS

Pressurized TBM tunnels are traditionally designed neglecting the pre-compression induced by the soil-structure interaction, largely due to the uncertainties involved in their estimation. Despite enormous progress in the monitoring equipments and technology, and also in the numerical modeling tools available, it is still a challenging task to estimate a lower-bound contact pressure induced in the tunnel lining, reliable enough for being employed in the joint bolt design.

A procedure for the risk assessment and the computation of the probability to have tension forces in the tunnel lining during the operation were presented and discussed in this paper.

The stochastic nature of some relevant parameters was acknowledged, and the calibration of probability density distributions was done based on experience, published data and usual correlations for selected critical parameters. In the same way, a statistical analysis of the river level and the properties of the concrete used to manufacture lining segments was detailed and used in the simulations. A first stage, consisting in a convergence analysis using 3D simulations of the excavation sequence was briefly discussed, and mean values for vertical crown displacement and effective stress were shown and used to define an equivalent ground relaxation for the 2D models.

After defining a reasonable range for $\Sigma M$ stage, a set of $6 \ x \ 2^6 = 384$ permutations were used in the 2D numerical model to estimate a range for the structural loads. A comparison with results obtained by the strain gauges during the tunnel construction in two rings was presented, showing a good agreement between the predictions and field measurements. Using this validation, the structural forces in service were estimated, concluding that the probability to support tension forces in this stage are negilgible for rings embedded in soft clay, while it increases for rings embedded in dense sands if the ground relaxation is not properly controlled.

## ACKNOWLEDGMENTS

The authors would like to acknowledge the Argentine Naval Hydrographic Service (www.hidro.gov.ar) for providing the water level time-history of the *Rio de la Plata* and Salini-Impregilo-Chediak for permission to publish the data contained in this paper.

## REFERENCES

Aradas, R., D. Tsingas, & M. Martini (2019). The design of a segmentally lined tunnel for a large sewer outfall. lote 3 - emisario planta riachuelo, argentina. In *World Tunnel Congress 2019, Naples - Italy*. International Tunnel Asociation.

CEB-FIP (1993). *MODEL CODE 1990*. Thomas Telford Publishing.

Ledesma, O. (2008). *Calibración del Cam Clay para suelos del post-pampeano*. Buenos Aires, Argentina: Tesis de grado de Ingeniería Civil - Universidad de Buenos Aires.

Muir Wood, A. (1975). The circular tunnel in elastic ground. *Géotechnique* 25(1), 115–127.

Sfriso, A. (1997). Formación post-pampeano: predicción de su comportamiento mecánico. *CLICJ - Caracas, Venezuela*, 1–10.